\documentclass[twocolumn,english,aps,prc,amsmath,amssymb,showpacs]{revtex4}
\usepackage[T1]{fontenc}
\usepackage[latin9]{inputenc}
\usepackage{amsmath}
\usepackage{amssymb}

\makeatletter

\providecommand{\tabularnewline}{\\}

\@ifundefined{textcolor}{}
{%
 \definecolor{BLACK}{gray}{0}
 \definecolor{WHITE}{gray}{1}
 \definecolor{RED}{rgb}{1,0,0}
 \definecolor{GREEN}{rgb}{0,1,0}
 \definecolor{BLUE}{rgb}{0,0,1}
 \definecolor{CYAN}{cmyk}{1,0,0,0}
 \definecolor{MAGENTA}{cmyk}{0,1,0,0}
 \definecolor{YELLOW}{cmyk}{0,0,1,0}
 }

\usepackage{babel}\usepackage{longtable}

\makeatother

\usepackage{babel}
\begin{document}

\title{The large-$j$ limit for certain 9-$j$ symbols---power law behaviour }

\author{L. Zamick}

\author{A. Escuderos}

\affiliation{Department of Physics and Astronomy, Rutgers University, Piscataway,
NJ 08854, USA}
\begin{abstract}
In a previous work, certain unitary 9-$j$ symbols were shown to go
asymptotically to zero in the large-$j$ limit. In this work, we examine
this in more detail and find an approximate power law for some unitary
9-$j$'s in the large-$j$ limit and exponentiial decrease for others
in this same limit. 
\end{abstract}
\maketitle

\section{Introduction}

A unitary 9-$j$ ($U9$-$j$) coefficient is related to a 9-$j$ symbol
via 
\begin{multline}
\langle(j_{1}j_{2})^{J_{12}}(j_{3}j_{4})^{J_{34}}|(j_{1}j_{3})^{J_{13}}(j_{2}j_{4})^{J_{24}}\rangle^{I}=\\
=[(2J_{12}+1)(2J_{34}+1)(2J_{13}+1)(2J_{24}+1)]^{1/2}\times\\
\times\begin{Bmatrix}j_{1} & j_{2} & J_{12}\\
j_{3} & j_{4} & J_{34}\\
J_{13} & J_{24} & I
\end{Bmatrix}
\end{multline}

We first note the well-known normalization relation 
\begin{equation}
\sum_{J_{13},J_{24}}\left|\langle(j_{1}j_{2})^{J_{12}}(j_{3}j_{4})^{J_{34}}|(j_{1}j_{3})^{J_{13}}(j_{2}j_{4})^{J_{24}}\rangle^{I}\right|^{2}=1~.
\end{equation}
 Another useful relation is: 
\begin{multline}
\sum_{J_{13},J_{24}}(-1)^{(J_{13}+J_{24})}\times\\
\times\left|\langle(j_{1}j_{2})^{J_{12}}(j_{1}j_{2})^{J_{34}}|(j_{1}j_{1})^{J_{13}}(j_{2}j_{2})^{J_{24}}\rangle^{I}\right|^{2}=0\label{eq:rel2}
\end{multline}
 if $J_{12}$ does not equal $J_{34}$. When $J_{12}$ is equal to
$J_{34}$, one gets $(-1)^{I}$.

\noindent In a previous work~\cite{ze13}, we noted that the {}``coupling''
$U9$-$j$ $\langle(jj)^{2j}(jj)^{2j}|(jj)^{2j}(jj)^{(2j-2)}\rangle^{I=2}$
decreased rapidly with increasing $j$ and went asymptotically to
zero. Indeeed the decrease is roughly exponential in $j$. This will
be discussed later.

The motivation for considering this class of $U9$-$j$'s was that
they enter into the overlap of the approximate wave functions for
two $I=2$ states. The components of these wave functions are the
following $U9$-$j$'s: \begin{subequations} \label{eq:wfab} 
\begin{align}
 & \langle(jj)^{2j}(jj)^{2j}|(jj)^{J_{p}}(jj)^{J_{n}}\rangle^{I=2}\label{eq:wfa}\\
 & \langle(jj)^{2j}(jj)^{(2j-2)}|(jj)^{J_{p}}(jj)^{J_{n}}\rangle^{I=2}\,.\label{eq:wfb}
\end{align}
 \end{subequations} It was found that the overlap was very small
and so~(\ref{eq:wfa}) and~(\ref{eq:wfb}) are good approximations
to the lowest two $I=2^{+}$ states when the E(9) interaction is used.
In this interaction for the $g_{9/2}$ shell, all two-body matrix
elements are set equal to zero except the one for $J=J_{\text{max}}=9$.

If there were no restriction on the integers $J_{p}$ and $J_{n}$,
(\ref{eq:wfa}) and (\ref{eq:wfb}) would be orthogonal. However,
we were considering a system of two protons and two neutrons in the
single $j$ shell. To satisfy the Pauli principle, $J_{p}$ and $J_{n}$
had to be even. As shown in Ref.~\cite{ze13}, the overlap, with
this restriction, is: $-1/2\langle(jj)^{2j}(jj)^{2j}|(jj)^{2j}(jj)^{(2j-2)}\rangle^{I=2}$.

The relation, generalized to any even total angular momentum $I$,
is 
\begin{multline}
\mathop{\sum_{\text{even}}}_{J_{p},J_{n}}\langle(jj)^{2j}(jj)^{2j}|(jj)^{J_{p}}(jj)^{J_{n}}\rangle^{I}\times\\
\times\langle(jj)^{2j}(jj)^{(2j-2)}|(jj)^{J_{p}}(jj)^{J_{n}}\rangle^{I}=\\
=-\frac{1}{2}\langle(jj)^{2j}(jj)^{2j}|(jj)^{2j}(jj)^{(2j-2)}\rangle^{I}\,.\label{eq:eveni}
\end{multline}

For odd total angular momentum $I$, the appropriate relation is 
\begin{multline}
\mathop{\sum_{\text{even}}}_{J_{p},J_{n}}\langle(jj)^{2j}(jj)^{(2j-1)}|(jj)^{J_{p}}(jj)^{J_{n}}\rangle^{I}\times\\
\times\langle(jj)^{2j}(jj)^{(2j-3)}|(jj)^{J_{p}}(jj)^{J_{n}}\rangle^{I}=\\
=-\frac{1}{2}\langle(jj)^{2j}(jj)^{(2j-1)}|(jj)^{2j}(jj)^{(2j-3)}\rangle^{I}\,.\label{eq:oddi}
\end{multline}

Regardless of the motivation, we here consider the behavior of the
$U9$-$j$'s $\langle(jj)^{2j}(jj)^{2j}|(jj)^{2j}(jj)^{(2j-2)}\rangle^{I}$
for even $I$ and $\langle(jj)^{2j}(jj)^{(2j-1)}|(jj)^{2j}(jj)^{(2j-3)}\rangle^{I}$
for odd $I$. For the sake of convenience, we will use the notation
$M_{j}^{\text{even}}(I)$ and $M_{j}^{\text{odd}}(I)$ for the latter
even-$I$ and odd-$I$ $U9$-$j$'s, respectively.

\section{Results}

\subsection{Even $I$}

We will here adopt a very simple approach. We just calculate a number
of $U9$-$j$'s and make reasonable guesses at the extrapolations.
We start by giving in Table~\ref{tab:mie} the values of $M_{j}^{\text{even}}(I)$
for the $g_{9/2}$ shell, i.e we consider $M_{9/2}^{\text{even}}(I)=\langle(\frac{9}{2}\frac{9}{2})^{9}(\frac{9}{2}\frac{9}{2})^{9}|(\frac{9}{2}\frac{9}{2})^{9}(\frac{9}{2}\frac{9}{2})^{7}\rangle^{I}$.

\begin{table}[htb]
 \caption{\label{tab:mie} Values of $M_{j}^{\text{even}}(I)$ (see text) for
all even total angular momenta $I$ in the $g_{9/2}$ shell.}

\begin{ruledtabular} %
\begin{tabular}{cr}
$I$  & \multicolumn{1}{c}{$M_{j}^{\text{even}}(I)$}\tabularnewline
\hline 
2  & $-0.000182$ \tabularnewline
4  & 0.000173 \tabularnewline
6  & $-0.000260$ \tabularnewline
8  & 0.000536 \tabularnewline
10  & $-0.001513$ \tabularnewline
12  & 0.006055 \tabularnewline
14  & $-0.037896$ \tabularnewline
16  & 0.491530 \tabularnewline
\end{tabular}\end{ruledtabular} 
\end{table}

A striking result is that all the $U9$-$j$'s are small except for
the one with the maximum value of $I$, namely $I=16$. A reasonable
speculation is that $M_{j}^{\text{even}}(I)$ will vanish in the limit
of large $j$ for all $I$ except $I_{\text{max}}=4j-2$. We can even
dare to speculate that the last one approaches a value of 1/2 in the
large-$j$ limit.

To test this in our simple approach, we go to a much higher $j$ shell:
$j=21/2$. The values of $M_{j}^{\text{even}}(I)$ for selected angular
momenta that we find are shown in Table~\ref{tab:j21}.

\begin{table}[htb]
 \caption{\label{tab:j21} Selected values of $M_{j}^{\text{even}}(I)$ (see
text) for the $j=21/2$ shell.}

\begin{ruledtabular} %
\begin{tabular}{cr}
$I$  & \multicolumn{1}{c}{$M_{j}^{\text{even}}(I)$}\tabularnewline
\hline 
2  & $-3.57861\times10^{-11}$ \tabularnewline
34  & $-8.35524\times10^{-5}$ \tabularnewline
36  & $9.27451\times10^{-4}$ \tabularnewline
38  & $-1.52261\times10^{-2}$ \tabularnewline
40  & \multicolumn{1}{c}{0.496870}\tabularnewline
\end{tabular}\end{ruledtabular} 
\end{table}

These results strongly support our speculations. The $I=2$ value
is now extremely small (of the order of $10^{-11}$) and all the others
are small except for $I=40$. The value for $I=40$ in the 21/2 shell
is closer to 1/2 than is the $I=16$ result in the 9/2 shell: 0.496870
vs. 0.491530.

The $U9$-$j$'s that go to a finite value in the large-$j$ limit
are said to exhibit classical behaviour and those that go to zero,
non-classical behaviour. Thus, we have only one $U9$-$j$ exhibiting
classical behaviour, the one with $I=2j+(2j-2)=4j-2$. We can gain
some insight into this behavior by noting that the relation of Eq.~(\ref{eq:eveni})
for the case $I=I_{\text{max}}=4j-2$ involves only one term: 
\begin{multline}
\langle(jj)^{2j}(jj)^{2j}|(jj)^{2j}(jj)^{(2j-2)}\rangle^{I_{\text{max}}}=\\
=-2\langle(jj)^{2j}(jj)^{2j}|(jj)^{(2j-1)}(jj)^{(2j-1)}\rangle^{I_{\text{max}}}\times\\
\times\langle(jj)^{2j}(jj)^{(2j-2)}|(jj)^{(2j-1)}(jj)^{(2j-1)}\rangle^{I_{\text{max}}}\,.
\end{multline}
 The first $U9$-$j$ on the right hand side approaches $1/\sqrt{2}$
in the large-$j$ limit. The second one approaches $-1/(2\sqrt{2})$
in the same limit.

\subsection{Odd $I$}

We can also consider odd total angular momentum. In that, we use Eq.~(\ref{eq:oddi}).
For $I=4j-3$ (the largest odd $I$), we again have that there is
only one term in the sum: 
\begin{multline}
\langle(jj)^{2j}(jj)^{(2j-1)}|(jj)^{2j}(jj)^{(2j-3)}\rangle^{(4j-3)}=\\
=-2\langle(jj)^{2j}(jj)^{(2j-1)}|(jj)^{(2j-1)}(jj)^{(2j-1)}\rangle^{(4j-3)}\times\\
\times\langle(jj)^{2j}(jj)^{(2j-3)}|(jj)^{(2j-1)}(jj)^{(2j-1)}\rangle^{(4j-3)}\,.\label{eq:loddi}
\end{multline}
 For $j=9/2$, the largest odd angular momentum is $I=15$, and the
left hand side of Eq.~(\ref{eq:loddi}) is equal to 0.42564827. We
expect that this will approach 1/2 in the large-$j$ limit.

Just as in the even-$I$ case, for odd $I$ less than 15, the values
are small and go to zero in the large-$j$ limit, as we can see in
Table~\ref{tab:mio}.

\begin{table}[htb]
 \caption{\label{tab:mio} Values of $M_{j}^{\text{odd}}(I)$ (see text) for
all odd total angular momenta $I$ in the $g_{9/2}$ shell.}

\begin{ruledtabular} %
\begin{tabular}{cr}
$I$  & \multicolumn{1}{c}{$M_{j}^{\text{odd}}(I)$}\tabularnewline
\hline 
3  & $1.410648\times10^{-3}$ \tabularnewline
5  & $-1.002940\times10^{-3}$ \tabularnewline
7  & $1.325594\times10^{-3}$ \tabularnewline
9  & $-2.673475\times10^{-3}$ \tabularnewline
11  & $8.145222\times10^{-3}$ \tabularnewline
13  & $-4.025313\times10^{-2}$ \tabularnewline
15  & \multicolumn{1}{c}{$0.425648$}\tabularnewline
\end{tabular}\end{ruledtabular} 
\end{table}

We briefly compare with $j=21/2$. The $U9$-$j$ is now $M_{21/2}^{\text{odd}}(I)=\langle(\frac{21}{2}\frac{21}{2})^{21}(\frac{21}{2}\frac{21}{2})^{20}|(\frac{21}{2}\frac{21}{2})^{21}(\frac{21}{2}\frac{21}{2})^{18}\rangle^{I}$.
We show in Table~\ref{tab:j21o} some relevant selected values.

\begin{table}[htb]
 \caption{\label{tab:j21o} Selected values of $M_{j}^{\text{odd}}(I)$ (see
text) for the $j=21/2$ shell.}

\begin{ruledtabular} %
\begin{tabular}{cr}
$I$  & \multicolumn{1}{c}{$M_{j}^{\text{odd}}(I)$}\tabularnewline
\hline 
15  & $9.084676\times10^{-10}$ \tabularnewline
35  & $9.407980\times10^{-4}$ \tabularnewline
37  & $-1.424514\times10^{-2}$ \tabularnewline
39  & \multicolumn{1}{c}{$0.430223$}\tabularnewline
\end{tabular}\end{ruledtabular} 
\end{table}

Thus, we see that the odd-$I$ case is similar to the even-$I$ case
in the sense that the $U9$-$j$ approaches zero in the large-$j$
limit for all $I$ except for the largest possible value $I=4j-3$.

\section{Power law behaviour near $I=I_{\text{max}}$}

\begin{table*}[htbp!]
 \caption{\label{tab:plb} Comparison of predicted versus calculated $U9$-$j$'s
for selected values of $I$ in the $j=41/2$ shell. We give also the
ratio of the predicted to the actual value for $n=1,2,3,4,5$.}

\begin{ruledtabular} %
\begin{tabular}{ccrrrc}
$I$  & $n$  & \multicolumn{1}{c}{$U(j=25/2)$} & \multicolumn{1}{c}{$U(j=41/2)$} & \multicolumn{1}{c}{$U_{\text{predicted}}(j=41/2)$} & Ratio \tabularnewline
\hline 
80  & 0  & \multicolumn{1}{c}{0.497390} & \multicolumn{1}{c}{0.498435} & \multicolumn{1}{c}{0.497390} & \tabularnewline
78  & 1  & $-0.126985\times10^{-1}$  & $-0.763220\times10^{-2}$  & $-0.774296\times10^{-2}$  & 1.0145 \tabularnewline
76  & 2  & $0.641513\times10^{-3}$  & $0.229283\times10^{-3}$  & $0.238516\times10^{-3}$  & 1.0405 \tabularnewline
74  & 3  & $-0.477053\times10^{-4}$  & $-0.100101\times10^{-4}$  & $-0.108152\times10^{-4}$  & 1.0804 \tabularnewline
72  & 4  & $0.471819\times10^{-5}$  & $0.589299\times10^{-6}$  & $0.652293\times10^{-6}$  & 1.1069 \tabularnewline
70  & 5  & $-0.579965\times10^{-6}$  & $-0.427933\times10^{-7}$  & $-0.488858\times10^{-7}$  & 1.1423 \tabularnewline
\end{tabular}\end{ruledtabular} 
\end{table*}

We again consider even $I$. We re-examine the $U9$-$j$ $\langle(jj)^{2j}(jj)^{2j}|(jj)^{2j}(jj)^{(2j-2)}\rangle^{I}$
close to $I=I_{\text{max}}$. We already noted that for $I=I_{\text{max}}$
this $U9$-$j$ increases asymptotically to 1/2 in the large-$j$
limit. We next consider $I=I_{\text{max}}-2$, $I_{\text{max}}-4$,
$I_{\text{max}}-6$, $I_{\text{max}}-8$, and $I_{\text{max}}-10$.
We find numerically that the first one goes slowly to zero approximately
as $1/j$, the second one as $1/j^{2}$, and so on. All these intriguing
and varying behavio$\times$rs deserve further study. We here give
the details.

We evaluate the selected $U9$-$j$'s for $j=25/2$ and $j=41/2$.
Then we use $j=25/2$ to predict what happens for $j=41/2$ via the
formula 
\begin{equation}
U_{\text{predicted}}(j=41/2)=U(j=25/2)\left(\frac{25}{41}\right)^{n}\,,
\end{equation}
 such that $I=I_{\text{max}}-2n$. We see in Table~\ref{tab:plb}
that there is close but not perfect agreement with the power law behaviour
$1/j^{n}$. The ratios predicted/actual for $j=41/2$ for $n=1,2,3,4,5$
are shown in the last column. The agreement is best for small $n$,
i.e. as $I$ gets closer to $I_{\text{max}}$.

\section{Exponential behaviour for I =2}

We now come back to the I=2 case. We again consider the U9-j symbol
U(j)= <(jj)$^{2j}$ (jj)$^{2j}$ | (jj)$^{2j}$ (jj)$^{(2j-2)}$>$^{I=2}$.
We speculate that the asymptic form is C e$^{(-aj)}$. to put this
to the test we consider 3 values - j, j+1 and j+2. if it were strickly
exponential then the ratio R1=U(j)/U(j+1) would be the same as the
ratio R2= U(j+1)/ U(j+2) . We present results for 3 values of j ,
40.5,60.5, and 100.5

j=40.5 , U(j)=-0.194664$\times$10$^{-45}$ , R1= 0.154314 $\times$10$^{2}$,
R2= 0.154445 $\times$10$^{2}$, R2-R1= 0.130816 $\times$10$^{-1}$

j=60.5 , U(j) =-0.292370 $\times$10$^{-69}$ , R1= 0.156142 $\times$10$^{2}$,
R2= 0.156203 $\times$10$^{2}$, R2-R1 = 0.605774 $\times$10$^{-2}$

j=100.5,U(j)= -0.426386 $\times$10$^{-117}$, R1= 0.157652 $\times$10$^{2}$,
R2= 0.157675 $\times$10$^{2}$ , R2-R1 = 0.231934 $\times$10$^{-2}$

We see that the difference R2-R1 gets smaller and smaller with increasing
j. The agreement is remarkable. For j=100.5 it is better than 1 part
in 6000.

We now try a more elaborate form C j$^{m}$e$^{(-\alpha j)}$ and
look for m. Consider the ratio R2/R1. For the functional form just
given it is equal to 

(j+1)$^{2m}$/( j (j+2))$^{m}$ For m=0 this has a value of one..
By comparing with the exact value of R2/R1 obtained from explicit
values of U9-j's we conclude that m=1.5. For example for j=40.5 the
values of R2/R1 is 1.00089299679 for m=1.5 whereas the exact value
is 1.0008477742. THe corresponding values for j=80.5 are 1.00022864342
and1.00022150836 .

Early works on 3-j, 6-j and 9-$j$ symbols were performed by atomic
and nuclear physicists, especially Wigner~\cite{w58} and Racah~\cite{r42}.
More recently, there have been extensive works by researchers in chemistry
and quantum gravity~\cite{aas08,hl10,yl11}. Very recently Van Isacker
and Macchiavelli {[}7{]} have considered the large j limit of shell
model matrix elements in the context of the problem of shears mechanisms
in nuclei. This involved 12-j symbols.In this work we have shown that
one can get a variety of behaviors of unitary 9-j's in the large j
limit. Some go to a finite value , some go to zero exponentially and
others go to zero via power laws i.e. 1/j$^{n}$ with varying n. We
have here found the reullts numerically. It will be a challenge to
get a better and perhaps more analytic understanding of these behaviors.

\end{document}